\newlength{\extralineskip}
\documentstyle[12pt]{article}
\begin{document}
\begin{titlepage}
\begin{flushright}
          \begin{minipage}[t]{12em}
          \large UAB--FT--361\\
                 March 1995
          \end{minipage}
\end{flushright}

\vspace{\fill}

\vspace{\fill}

\begin{center}
\baselineskip=2.5em

{\LARGE  ON A LIGHT SPINLESS PARTICLE  \\
COUPLED TO PHOTONS}
\end{center}

\vspace{\fill}

\begin{center}
{\sc Eduard Mass\'o and Ramon Toldr\`a}\\

     Grup de F\'\i sica Te\`orica and Institut de F\'\i sica
     d'Altes Energies\\
     Universitat Aut\`onoma de Barcelona\\
     08193 Bellaterra, Barcelona, Spain
\end{center}

\vspace{\fill}
\begin{center}

\large ABSTRACT
\end{center}
\begin{center}
\begin{minipage}[t]{36em}

A pseudoscalar or scalar particle $\phi$ that couples to two
photons but not
to leptons, quarks and nucleons would have effects in most of the
experiments searching for axions, since these are based on the
$a \gamma \gamma $ coupling. We examine the laboratory,
astrophysical and
cosmological constraints on $\phi$ and study whether it may
constitute a substantial part of the dark matter. We also
generalize the
$\phi$ interactions to possess $SU(2) \times U(1)$ gauge
invariance, and analyze the phenomenological implications.

\end{minipage}
\end{center}
\vspace{\fill}

\end{titlepage}

\clearpage
\addtolength{\baselineskip}{\extralineskip}

\section{Introduction and Motivation}

The Peccei-Quinn symmetry \cite{Peccei/Quinn} is still the most
attractive
solution to the strong CP problem of QCD. As a consequence of the
spontaneous
breaking of that symmetry, the axion is born \cite{Weinberg...}.
The axion properties and their phenomenological consequences have
been studied in depth ( for a review see \cite{Peccei/Jarlskog} ),
and some experiments trying to discover the axion are under way
( for a review see \cite{Sikivie} ). Axions might be constituents
of the dark mass of the Universe, and this makes the search
experiments even more fascinating.

Almost all experiments so far designed to search for light axions
make use of the coupling of the axion to two photons
\begin{equation} \label{lag.axions}
{\cal L} = \frac{1}{8} \ g_{a \gamma \gamma} \
           \varepsilon_{\mu \nu \alpha \beta}
           F^{\mu \nu} F^{\alpha \beta} \ a
\end{equation}

The coupling $g_{a\gamma\gamma}$ is proportional to the axion
mass $m_a$
\begin{equation} \label{axioncoupling}
g_{a \gamma \gamma} \approx \frac{\alpha}{2\pi} \
                            \frac{m_a}{1 \ \mbox{eV}} \
                            10^{-7} \ \mbox{GeV}^{-1}
\end{equation}

An interesting question is whether these dedicated experiments
are 1) only
sensible to the axion or 2) they could discover another class of
particles. The answer is 2). Indeed, any light pseudoscalar
particle $\phi$ coupled to two photons
\begin{equation} \label{lag.pseu}
{\cal L} = \frac{1}{8} \ g \ \varepsilon_{\mu \nu \alpha \beta}
                           F^{\mu \nu} F^{\alpha \beta} \ \phi
\end{equation}
with a strong enough coupling $g$ would induce a positive
signal in some of the axion searches. Of course, a scalar
particle coupled to two photons would also be
detected in such experiments. To simplify the presentation of
the paper, first we will thoroughly discuss the pseudoscalar
case. In Sec. 6 we will compare the scalar to the pseudoscalar
case.

With all this in mind, we have studied the phenomenology and
consequences of a light particle $\phi$ that couples
{\bf only} to two photons with strength $g$.

We consider exclusively this type of interaction, Eq.
(\ref{lag.pseu}), since it is the existence of this
interaction the only requirement for having a signal in the
axion experiments. By making this assumption, however, we
are not
generalizing the axion. Our particle $\phi$ cannot be
identified with the
axion, since the axion couples to leptons, quarks and
nucleons, and $\phi$
does not. In this spirit, we will also assume that the
coupling $g$ and the
mass $m$ of the $\phi$ particle are not related, as they
are for the axion,
Eq. (\ref{axioncoupling}). In principle, we should consider
arbitrary
$\phi$ masses, but since we know that axion experiments are
sensitive to very light axions, we will restrict the range
of masses; we will only consider $m \leq 1$ GeV.

In this paper, we will investigate the laboratory,
astrophysical and
cosmological constraints on $\phi$. Some of the axion
constraints can be
directly translated into constraints on $g$ and $m$, but
some cannot. We will
also answer the question whether the relic $\phi$ particles
can be, for some
range of parameters, the dark matter of the Universe. Another
issue we will
study is the consequence of adding other couplings to
$\gamma \gamma \phi$
in such a way that the full $SU(2) \times U(1)$ gauge invariance
holds at high energies.

We finish this section with some general remarks. As we said,
the motivation that has led us to assume a light particle coupled
only
to photons is the fact that experiments are sensitive to such a
possibility.
As far as we know, there is not a current theoretical model where
such
peculiarities arise. In fact, one may even wonder whether it can
ever occur.
The point is that we know that the coupling of (quasi) Goldstone
bosons to
photons proceeds through anomalous triangle graphs, where the
boson couples
to charged particles. This is the situation for the neutral pion
and in the
axion model. One may argue that in order to couple $\phi$ to
photons,
$\phi$ has to couple to charged particles, and one may conclude
that our assumption of absence of couplings to matter is
inconsistent.

We would like to point out that one may think of scenarios where
the only
coupling that one may constrain at low energies is $g$ in
Eq. (\ref{lag.pseu}). We need to introduce
particles that are very heavy and carry a new quantum number. We
also have to
impose that $\phi$ carries also this quantum number, and that the
known leptons
and quarks do not. The anomalous graphs with a triangle loop of
new particles
would then induce the effective coupling of Eq. (\ref{lag.pseu}).
For heavy enough new particles, the important coupling of $\phi$
at low energies would be to photons, and all the
constraints discussed in this paper do not need to be modified or
reconsidered.

A related point is the fact that the effective Lagrangian
(\ref{lag.pseu})
can only be used for energies $E \ll g^{-1}$. We keep in mind this
restriction in all the calculations.


\section{Laboratory constraints: non-dedicated experiments}

In the next section we will discuss the consequences that the
experiments
designed to look for axions have for our $\phi$ particle. Before
that
discussion, we show in this section that there are other laboratory
experiments
that also give limits on the parameters of $\phi$: its mass $m$ and
coupling $g$.

The quarkonium data can be used to constrain $g$. The most
restrictive limit comes from the absence of the decay
\begin{equation} \label{upsilon.decay}
\Upsilon (1S)\longrightarrow \phi \ \gamma
\end{equation}
where $\phi$ does not decay in the detector. In Fig. 1(a) we show
the diagram
giving rise to that decay, and we would like to stress that it is
{\bf not} the
same diagram used to constrain the axion properties in quarkonium
decays
$\Upsilon (1S)\rightarrow a \ \gamma$ \cite{Weinberg...}. It is
easy to find the branching ratio corresponding to the exotic decay
(\ref{upsilon.decay})
\begin{equation} \label{BRupsilon}
BR(\Upsilon (1S)\rightarrow \phi \ \gamma ) = \frac{1}{8\pi \alpha}
             \  g^2 m^{2}_b \ BR(\Upsilon (1S)\rightarrow e^{+}e^-)
\end{equation}
where we have used that $m_{b}\gg m$. In order to use the
experimental data, we need to impose that $g$ is small enough in
such a way that $\phi $ does not decay inside the detector
\begin{equation} \label{length}
\frac{p_{\phi}}{m} \tau > L_{det}
\end{equation}
where the lifetime $\tau$ is given by
\begin{equation} \label{lifetime}
\tau = \frac{64\pi}{g^2 m^3}
\end{equation}

Using the Crystal Ball data \cite{Ball} we find the limit
\begin{equation}
g \leq 4.2 \ 10^{-3} \ \mbox{GeV}^{-1}
\end{equation}
The forbidden region is shown in Fig. 2. We set $L_{det} = 2$ m
as a rough estimate; our final result does not strongly depend
on the precise value of $L_{det}$.

Positronium decay into a single photon plus invisible particles
leads
to limits that are less restrictive than quarkonium decays.
The decay
$K^{+}\rightarrow \pi^{+}+missing$ and similar were used to
constrain axion
decays \cite{Peccei/Jarlskog} but do not restrict the coupling
$g$ of the $\phi$ particle.

We find however another high energy process from which more
stringent limits
can be extracted. The process is
\begin{equation} \label{annihilation}
e^{+}e^- \longrightarrow \phi \ \gamma
\end{equation}
and the Feynman diagram is shown in Fig. 1(b). The cross
section in the
limit $m,m_{e}\ll \sqrt{s}$ is
\begin{equation} \label{cross.section}
\sigma(e^{+}e^- \rightarrow \phi \ \gamma) = \frac{\alpha}{24}
                                       \ g^2 \ f(\theta_{min})
\end{equation}
where $f(\theta_{min})$ is a factor less than one determined
by the angular resolution of the experimental device.

The signature of the process (\ref{annihilation}) is a single
photon
observed in the final state, with unbalanced momentum. The
standard
model process $e^{+}e^- \rightarrow \nu \bar{\nu} \gamma$,
with $\nu =
\nu_{e},\nu_{\mu}$ or $\nu_{\tau}$, has identical signature. Some
experiments \cite{Hearty,Others} find a signal consistent with
the standard
model contribution, and therefore they are able to place bounds
on the
anomalous single-photon production. We use this fact to determine
a forbidden region in the $g$ versus $m$ plot. In fact, it is
the ASP
results \cite{Hearty} that lead to the most restrictive limits
on the $\phi$ parameters
\begin{equation}
g \leq 5.5 \ 10^{-4} \ \mbox{GeV}^{-1}
\end{equation}
provided $m \ll 29$ GeV.
This is what is shown in Fig. 2.


\section{Laboratory constraints: dedicated experiments}

In the last past few years, some dedicated experiments have been
designed
to search for axions. Many of these experiments make use of the
axion coupling to two photons: laser experiments
\cite{Maiani,VanBibber1,Cameron},
solar flux detection \cite{Sikivie.idea,VanBibber2,Lazarus},
telescope search \cite{Kephart/Turner,Bershady}
and microwave cavity experiments \cite{Wuensch/Hagmann}. We will
find that the first three types of experiments
lead to constraints on the
properties of $\phi$, but not the last one. We will
discuss the four
types of experiments in turn. In the discussion we will see that
some experiments need astrophysical
or cosmological assumptions
in order to extract consequences from their observations.

The laser experiments consist in the study of laser beam
propagation
through a transverse magnetic field \cite{Maiani,Cameron}.
The production
of real $\phi$ particles, as in Fig. 3(a), would produce a rotation
of the beam polarization, while the emission and absorption of a
virtual
$\phi$, as in Fig. 3(b), would contribute to the ellipticity of
the laser beam. Such effects are not observed and their absence
implies the following constraint \cite{Cameron}
\begin{equation} \label{Cam}
g \leq 3.6 \ 10^{-7} \ \mbox{GeV}^{-1}
\end{equation}
This limit is valid provided $m \leq 1$ meV, which is the
condition for coherent $\phi$ production.

A slightly less restrictive limit is obtained when considering
the photon regeneration effect that would also occur in the
presence of a coupling of $\phi$ to two photons
\cite{VanBibber1,Cameron}( see Fig. 3(c) ).

The laser experiment limits are shown in Fig. 2. The limits
on $g$
based on optical techniques are expected to improve by a factor
of about
40 when the PVLAS experiment \cite{Bakalov} will take and analyze
data.
We would like to stress that the constraints from these laser
experiments do not depend on astrophysical or cosmological
assumptions.

The search for solar axions \cite{VanBibber2,Lazarus} is based
on Sikivie's idea \cite{Sikivie.idea}
that axions produced in the Sun can be converted into
X-rays in a
static magnetic field. One can use the absence of such signal
to place a
limit on the coupling $g$ of $\phi$ to two photons. The Sun
produces
$\phi$ particles through the Primakoff effect
( see Fig. 4(a) ), and these particles reconvert to X-rays by
means of the inverse process shown in Fig. 4(b). The limit
is \cite{Lazarus}
\begin{equation}  \label{Laz}
g \leq 3.6 \ 10^{-9} \ \mbox{GeV}^{-1}
\end{equation}
Again, this limit is only true when the conversion process is
coherent
in the magnetic telescope. Coherence is preserved for
$m \leq 0.03$ eV.
When $0.03$ eV $\leq m \leq 0.11$ eV a slightly less
restrictive limit is
obtained \cite{Lazarus}. The constraints on the $\phi$
properties
obtained from the solar flux experiment are plotted in Fig. 2.
To reach the numerical limit on $g$, Eq. (\ref{Laz}), one
relies on the calculated axion luminosity from the
Sun \cite{Raffelt}.

The axion search program in BINP
( Novosibirsk ) will use a similar procedure to probe
the coupling $g$ \cite{Vorobyov}.

We turn now our attention to the telescope search experiment
\cite{Kephart/Turner,Bershady}, which was motivated by the
axion window in the eV
range \cite{Kolb}. The unsuccessful search can be used to
constrain
the properties of $\phi$ when $m$ is also in the eV range.
The assumptions one needs are reasonable \cite{Kephart/Turner}:
$\phi$
particles have been in thermal equilibrium in the early Universe due
to their photon interaction, leaving a relic density after
decoupling.
These cosmological $\phi$ particles will be found in galactic
clusters, since they fall with baryons into the potential wells
that arose in the evolution of the early Universe. The decay
$\phi \rightarrow \gamma \gamma$, provided $\phi$ has a mass in
the range
3 eV to 8 eV, should produce detectable lines when observing rich
clusters of galaxies. Bershady, Ressell and Turner
\cite{Bershady} have reported a null signal for such line of
radiation, which exclude the region indicated in our Fig. 2.

We finally consider the microwave cavity experiments
\cite{Wuensch/Hagmann},
which search for the signal of the conversion of halo axions, with
$E\sim 10^{-5}$ eV, in a magnetic field following
the ideas of Ref.\cite{Sikivie.idea}( see Fig. 4(b) ).
The search,
up to now, has been unsuccessful. To get strong limits on
the coupling
of axions to photons, it is assumed that axions form the dark
galactic
halo. While this assumption is tenable in the case of axions,
it is far from being realistic when considering $\phi$, as we
now discuss.

For the range of $g$ and of $m$ of interest in microwave
experiments, relic $\phi$ are relativistic. As we will see
in Sec. 5, today's temperature of $\phi$ is bounded by
$T_{\phi} \leq 10^{-4}$ eV. It follows that the contribution
of $\phi$ to the total energy of the Universe, normalized
to the critical density, is
\begin{equation} \label{Omega}
\Omega_{\phi} = \frac{\rho_{\phi}}{\rho_{crit.}} \leq 10^{-5}
\end{equation}
{\it i.e.}, the abundance of $\phi$ particles fails by several
orders of magnitude to account for the halo dark matter. We
conclude that the microwave experiments give very poor
constraints on the properties of $\phi$.

\section{Astrophysical constraints}

Stellar interiors can produce $\phi$ particles - provided $\phi$ is
light
enough - and these particles can escape almost freely - provided
$\phi$
interacts weakly enough with radiation. This effect is a potentially
important energy loss mechanism. The confrontation
of astrophysical observations with the modifications caused by
non-standard
energy losses on star evolution leads to astrophysical constraints
on the $\phi$ properties.

In the case of the Sun and red giant evolution it turns out that
the detailed
studies that have been done in the literature for the axion can
be used
without changes to get limits on the coupling $g$ and the mass $m$ of
$\phi$. To be more specific, the so-called hadronic or KSVZ axion is
dominantly
produced in the Sun and in a red giant interior through the Primakoff
process. It is also through this process that $\phi$ is produced, so
that we can use the axion limits. We will borrow the numerical
results found by Raffelt and Dearborn \cite{Raffelt/Dearborn}.
The reader interested in more details can consult Raffelt's review
\cite{Raffelt.PR}, where a full list of references can be found.

To get a rigorous limit on $g$ from the Sun we use the numerical
studies of \cite{Raffelt/Dearborn}, that follow the evolution of
a star
with $M=M_{\odot}$ that loses energy with the emission of light
particles produced by the Primakoff effect. For reasonable values of
the presolar helium concentration the present luminosity $L$ of the
star corresponds to the Sun luminosity $L=L_{\odot}$ as long as
\begin{equation} \label{Sun g}
g \leq 2.5 \ 10^{-9} \ \mbox{GeV}^{-1}
\end{equation}
The limit is valid when the $\phi$ mass is less than the central
temperature of the Sun, $T\approx 1$ keV.

When $m$ is larger than the solar core temperature one has
to take into account that the number density of photons
with enough energy to produce particles $\phi$ quickly
decreases when $m$ increases. Therefore, for 1 keV $\leq m
\leq$ 60 keV, the corresponding bound on $g$ is less
stringent than for smaller masses, and for masses larger
than 60 keV there is no bound. This effect can be seen
in our results plotted in Fig. 2.

Also, when $g$ is large enough, $\phi$ particles are trapped
in the Sun mainly due to their decay into two photons if
$m$ is larger than a few keVs, or to rescattering through
the inverse Primakoff effect if $m$ is smaller. To bound
$g$, we have calculated the effective mean opacity
of $\phi$ and compared it with the photon mean opacity, along
the lines described in Ref. \cite{Starkman/Salati}. It turns
out that $\phi$ would be allowed for quite a strong coupling
$g$, and in fact much of the excluded region in this trapping
regime was already ruled out by the laboratory experiments
presented in section 2; when there is such overlap we do not
show the astrophysical bound in Fig. 2.

Raffelt and Dearborn \cite{Raffelt/Dearborn} have also followed the
evolution of stars with properties similar to the stars of the
open cluster M67, paying attention to plasmon mass, screening
and degeneracy effects. The helium burning lifetime is shortened
when $\phi$
is emitted from the stellar interior. Star count in M67 and other
open clusters would be in sharp conflict with the effect of energy
loss in form of $\phi$ particles unless
\begin{equation} \label{Helium.g}
g \leq 1\ 10^{-10} \ \mbox{GeV}^{-1}
\end{equation}
for masses less than $T\approx 10$ keV. For 10 keV$\leq m \leq$
300 keV, the suppression in the number density of photons with
enough energy to produce $\phi$ particles makes the
bound on $g$ less and less stringent until it disappears.

Also, $\phi$ would be trapped in the red giant interiors for
a relatively
large value of the coupling $g$. Again, we use the procedure
presented in Ref. \cite{Starkman/Salati} to limit $g$, as we
did in the solar case.
In Fig. 2 we display these limits when there is not overlapping
with the limits coming from laboratory experiments.

Finally, we will consider the constraints coming from the
observation
of the neutrino pulse from the supernova SN1987A. The limits
on $g$
are less stringent than the ones obtained from the Sun and
red giants,
and for this reason an estimation of the limit will be enough for
our purposes. Let us start assuming that $m$ is smaller than
the temperature of the supernova core.
In this core, $\phi$ production is
mainly due to the processes ( see Fig. 5 )
\begin{eqnarray} \label{proc.SN}
e \gamma & \longrightarrow & e \phi \\
p \gamma & \longrightarrow & p \phi \\
p n & \longrightarrow & p n \gamma \phi
\end{eqnarray}
We stress that these are {\bf not} the main processes for axion
production in supernovae.

The cross sections for these processes can be estimated as follows:
\begin{eqnarray} \label{cross.SN}
\sigma (e \gamma \rightarrow e \phi) & \approx &
\sigma (p \gamma \rightarrow p \phi) \\
& \approx & \alpha \ g^2 \\
\sigma (p n \rightarrow p n \gamma \phi) & \approx &
                       \frac{\alpha}{\pi^3}
                       \ \sigma (p n \rightarrow p n)\ g^2 \ T^2
\end{eqnarray}
where $T \approx 50$ MeV is the temperature of the supernova
core and $\sigma (p n \rightarrow p n) \simeq 100$ mb is the cross
section for the process $p n \rightarrow p n$.
These cross sections can be used in turn to estimate the rate of
energy drain:
\begin{eqnarray} \label{drain.SN}
\dot{E}(e \gamma \rightarrow e \phi) & \approx &
\dot{E}(p \gamma \rightarrow p \phi) \\
    & \approx & V \ n_{p} \ n_{\gamma}
    \ \sigma (p \gamma \rightarrow  p \phi) \ T \\
\dot{E}(p n \rightarrow p n \gamma \phi) & \approx & V \ n_p \ n_n \
    \sigma (p n \rightarrow p n \gamma \phi) \ \sqrt{\frac{T}{m_N}}
    \ T \\
\end{eqnarray}
Here $V\simeq 4\ 10^{18}$ cm$^3$ stands for the volume of the
supernova core and $m_N$ is the nucleon mass. The number densities
we will adopt are $n_n \simeq 7\ 10^{38}$ cm$^{-3}$ and $n_p
\simeq n_e
\simeq 3\ 10^{38}$ cm$^{-3}$. The constraint from the neutrino
signal emitted by SN1987A is \cite{Kolb,Raffelt.PR,Schramm}
$\dot{E} \leq 10^{52}$ erg/sec, or in terms of $g$,
\begin{equation} \label{SN.g.1}
g \leq 10^{-9} \ \mbox{GeV}^{-1}
\end{equation}
valid for $m\leq$ 50 MeV.
As we already said, the SN1987A limit is less stringent than other
astrophysical limits. However, the range of $m$ for which the limit
holds is larger. For 50 MeV $\leq m \leq$ 500 MeV this bound
becomes smaller due to the low number of photons with enough
energy to produce $\phi$ particles.

To find the limits in the trapping regime, we follow Ref.
\cite{Starkman/Salati} to calculate the effective $\phi$
opacity $\kappa_{\phi}$, dominated by decay, and impose
\cite{Shapiro}
\begin{equation} \label{kappa}
\kappa_{\phi} > \kappa_{\nu} \simeq 8\
10^{-17} \mbox{cm}^2/\mbox{g}
\end{equation}

All these constraints are shown in Fig. 2.

One can use data on pulsar signals to probe the $g$ coupling,
provided the mass is very small, $m\leq 10^{-10}$ eV
\cite{Mohanty}. Magnetic fields of pulsars create a $\phi$
background, and pulsar signals propagating through this
background show a time lag between different modes of
polarization. The limit from present data is
$g\leq 2\ 10^{-11}$ GeV$^{-1}$.

A similar limit, $g\leq 2.5\ 10^{-11}$ GeV$^{-1}$, also
valid for small masses, $m\leq 10^{-9}$ eV, has been
obtained by Carlson \cite{Carlson}. He has studied
X-ray conversion of $\phi$ produced in stellar cores
and found the limit on $g$ using HEAO1 satellite data
on $\alpha$-Ori X-ray emission.


\section{Cosmology of $\phi$}

In this section, we will study the cosmological evolution of
the $\phi$
species until the present time. We will discuss to which extent
$\phi$
can be dark matter of the Universe, and also the constraints from
primordial nucleosynthesis on the $\phi$ parameters.

In the early Universe, $\phi$ is in equilibrium due to the processes
\begin{eqnarray} \label{eq.proc}
A \phi & \longleftrightarrow & A \gamma
          \; \; \; \;\mbox{and changing } A \; \mbox{by }  \bar{A}
          \nonumber \\
\gamma \phi & \longleftrightarrow & A \bar{A}.
\end{eqnarray}
Here $A$ stands for any particle with electric charge
$Q_A \neq 0$ present
in the primordial plasma. The interaction rate can be written as
\begin{equation} \label{Gamma}
\Gamma \simeq \frac{\alpha}{24} \ g^2 \ n
\end{equation}
$n$ is given by
\begin{equation}
n \simeq \frac{\zeta (3)}{\pi^2}\ f(T) \ T^3
\end{equation}
where
\begin{equation} \label{f(T)}
f(T) \approx   \sum_{A} g_A Q_A^2
\end{equation}
and $g_A$ are related to the internal degrees of freedom of $A$.
In this formula the sum runs over all the charged particles with
$m_A<T$. We are making the usual approximation of considering only
the relativistic degrees of freedom, whose contribution is dominant
( the contribution of the non-relativistic degrees of freedom is
exponentially suppressed).

The expansion of the Universe is characterized by the rate
\begin{equation} \label{Hubble}
H = 1.66 \ \sqrt{g_*(T)} \ \frac{T^2}{m_{Pl}}
\end{equation}
where $g_*(T)$ are the relativistic degrees of freedom contributing
to the energy density - we follow the notation of \cite{Kolb}. The
$\phi$ species decouples when both rates meet
\begin{equation} \label{decoupling}
H(T_f) = \Gamma (T_f)
\end{equation}
The freeze out temperature is a ( decreasing ) function of $g$,
and also depends on $m$. It can be calculated numerically,
but before
we do it, we present two general peculiarities that will help to
understand the cosmological evolution of $\phi$.

First, we can show that $T_f$ is bounded by $T_f \geq 0.2$ MeV.
Using that
for $T_f \leq 0.2$ MeV the only particle contributing
to the processes
(\ref{eq.proc}) is $A = e$, that $n_e \sim 10^{-10}\ T^3$ and that
$g_* \simeq 3.4$ we can deduce from (\ref{decoupling})
\begin{equation}
T_f \sim 0.2 \
        \left(\frac{0.04\ \mbox{GeV}^{-1}}{g} \right)^2 \mbox{MeV}.
\end{equation}
Since laboratory experiments exclude $g\geq 0.04$ GeV$^{-1}$ for
$m\leq 10$ MeV, we conclude that $T_f \geq 0.2$ MeV. The $\phi$
freeze out occurs before $e^+e^-$ annihilation.

Second, we can show that $\phi$ can only be a hot relic.
Writing $\Gamma$
as a function of the $\phi$ lifetime $\tau$, and $H = (2t)^{-1}$
( RD Universe
and we set $\Omega = 1$, but other values of $\Omega$ lead
to the same
conclusion ), the freeze-out condition (\ref{decoupling}) can be
written as
\begin{equation} \label{hot relic 1}
\frac{3T_f}{m} \approx \left(\frac{\tau}{f(T_f)\ t_f}\right)^{1/3}.
\end{equation}
We now force that $\tau \geq t_0 \sim 10^{17}$ sec,
since we want $\phi$
to survive until the present time. Also, since $T_f \geq 0.2$ MeV,
we have that $t_f \leq 1$ sec. It follows that
\begin{equation} \label{hot relic 2}
\frac{3T_f}{m} \geq
               \left(\frac{10^{17}}{f(T_f)}\right)^{1/3} \gg 1,
\end{equation}
{\it i.e.} if $\phi$ is a relic, it is a hot relic.
(These conclusions are based on the processes (\ref{eq.proc}). One
can show that the process $\phi \leftrightarrow \gamma \gamma$
does not modify them.)

After showing these two general features, we turn our
attention to
the $\phi$ abundance. Specifically, we would like to
elaborate on the
question whether $\phi$ can be a substantial component of the dark
matter of the Universe.

If $\phi$ is today a relativistic particle, it has a very
small abundance. The density would be less than the relic photon
density
\begin{equation} \label{relativistic}
\Omega_{\phi} h^2 < \Omega_{\gamma} h^2 \simeq 2.6\ 10^{-5}.
\end{equation}
Thus we will concentrate on the case that $\phi$ is
a non-relativistic
particle, $m > O(10^{-4}\ \mbox{eV})$. The abundance is
\begin{equation} \label{relic}
\Omega_{\phi} h^2 = 7.8 \ 10^{-2} \frac{m(\mbox{eV})}{g_{*S}(T_f)}
\end{equation}
where $g_{*S}(T)$ are the effective degrees of freedom that
contribute to the entropy of the Universe at a
temperature $T$ \cite{Kolb}.
The calculation of $\Omega_{\phi}$ when 0.2 MeV
$\leq T_f \leq$ 300 GeV
is straightforward since we know the functions $g_{*}(T)$,
$g_{*S}(T)$
\cite{Olive} and $f(T)$. We first calculate $T_f$ solving Eq.
(\ref{decoupling}) and then we introduce $g_{*S}(T)$ in Eq.
(\ref{relic})
to get the $\phi$ contribution to the energy density
of the Universe.
Freeze out at $T_f \sim 300$ GeV corresponds to a coupling
$g \sim 10^{-9}$
GeV$^{-1}$. We would like to know the abundances for smaller
values of $g$,
say until $g \sim 10^{-12}$, that would correspond to a freeze-out
temperature $T_f \leq 10^9$ GeV. We have to extrapolate
$g_{*}(T)$, $g_{*S}(T)$
and $f(T)$ for $10^9 \geq T \geq 300$ GeV. We will consider three
plausible scenarios:
\begin{itemize}
\item (A). The SM desert: there are no particles with masses
above $\sim 300$ GeV so that $g_{*}(T)$, $g_{*S}(T)$ and $f(T)$ are
constant at high energies.
\item (B). The MSSM desert: one finds the SUSY partners at
$\sim 300$ GeV
so that the functions $g_{*}(T)$, $g_{*S}(T)$ and $f(T)$
increase in one
step, and stay flat at high temperatures.
\item (C). Power law extrapolation: the function $g_{*}(T)$
for 200 MeV
$\leq T \leq 300$ GeV can be approximated by a power-law function,
$g_{*}(T) \sim T^{0.1}$.
We can extrapolate $g_{*}(T)$ at high temperatures
by letting it to increase according to this power-law function.
The functions
$g_{*S}(T)$ and $f(T)$ are extrapolated in a similar way.
\end{itemize}

It should be clear that we ignore how to extrapolate $g_{*}(T)$,
since that depends on the unknown new physics
and new particles above the weak scale. We will in turn consider
the three
scenarios described above, calculate $\Omega_{\phi}$ and see the
consequences.
In a sense scenarios (A) and (C) represent two extreme possibilities:
$g_{*}(T)$
does not increase at all in (A), while it increases at high
temperatures at the
same pace as it does at low temperatures in (C). The case (B)
is a model motivated scenario.
In Fig. 6 we present our results, in the form of lines of
constant $\Omega_{\phi}h^2$. Two possibilities are obviously
interesting:
$\phi$ being the cosmological dark matter of a critical Universe
or $\phi$
being only the dark matter in galactic halos. For the first
possibility
to make sense we should have $\Omega_{\phi} = 1$; for the
second a necessary
condition is $\Omega_{\phi} \geq 0.02h^{-1}$.
Since $0.5 \leq h \leq 1.0$, the
interesting range is $0.01 \leq \Omega_{\phi}h^2 \leq 1$.
The two extremes
of the range are plotted in Fig. 6, where also the requirement that
$\phi$ has survived until the present time, $\tau > H_{0}^{-1}$, is
plotted\footnote{The requirement should be $\tau > t_{Univ.}$.
The lifetime of
the Universe $t_{Univ.}$ is proportional to $H_{0}^{-1}$ with the
proportionality factor depending on $\Omega$. Also, $H_{0}$ is
known up
to a factor of 2. Our conclusions do not depend on these details.}.
Both
conditions $\Omega_{\phi}h^2 = 1$ and $\Omega_{\phi}h^2 = 0.01$
are worked
out and plotted for each of the three scenarios (A), (B) and (C)
presented above.

The dotted region in Fig. 6 is the range of couplings and
masses such that $\phi$ is the dark matter. We see that
$\phi$ must have
at least a mass $m \geq 10$ eV. The maximum mass that is
interesting for
dark matter depends on how small we allow $g$ to be. Let us
impose that
$g \geq 10^{-12}$ GeV$^{-1}$. Then we have $m \leq 10$ keV. For
10 eV $\leq m \leq 100$ eV, $\phi$ could be the
galactic dark matter;
for 1 keV $\leq m \leq 10$ keV $\phi$ is the cosmological
dark matter.

The interesting $\phi$ masses for dark matter
are unfortunately away
from the mass range where undergoing experiments could be able to
detect $\phi$. As we mention in Sec. 3, laser experiments work
for $m \leq 10^{-3}$ eV, solar axion detection is possible for
$m \leq 10^{-1}$ eV and microwave experiments are restricted to
$E \sim 10^{-5}$ eV. It is clear that these experiments are not
sensitive to a particle $\phi$ coupling only to two photons and
constituting the dark matter. The telescope search experiment has
3 eV $\leq m \leq 8$ eV, which is near the mass range relevant
for dark matter. Since the calculation of the $\phi$ abundances
has some inherent uncertainties, we conclude that the telescope
search may be sensitive to a $\phi$ that forms the galactic
dark halo.

A last point relevant for $\phi$ being dark matter
is the fact that
$\phi$ would be hot dark matter, as it was shown at the beginning
of this section. Formation of the large structures of the Universe
seems however to require a large proportion of cold dark matter,
so that $\phi$ would not have a role in structure formation.

We finally discuss the constraint on the $\phi$
parameters that arises
when considering the big bang nucleosynthesis ( BBN ) of light
elements.
The comparison of the observed primordial abundances
with the predictions
of the standard model of the early Universe places a bound
on the expansion
rate at $T \approx 1$ MeV. In terms of relativistic
degrees of freedom this bound reads \cite{Walker}
\begin{equation} \label{experiment}
\Delta g_{*}(T=1 \ \mbox{MeV}) \leq 0.5
\end{equation}

The contribution of $\phi$ to the effective degrees
of freedom at the
nucleosynthesis epoch, provided $m \ll 1$ MeV, is given by
\begin{equation} \label{phi role}
\Delta g_{*}^{\phi}(T) = \left(\frac{T_{\phi}}{T}\right)^4
\end{equation}
where $T \simeq 1$ MeV. We allow for the case that
after the $\phi$ decoupling
and before the nucleosynthesis epoch there has been pair
annihilations to
photons, in such a way that at $T \simeq 1$ MeV one
has $T_{\phi} \neq T$.
How different the two temperatures are is a function
of $T_d$, the decoupling
temperature. Entropy conservation gives the
relation \cite{Kolb,Olive}
\begin{equation} \label{T phi}
\frac{T_{\phi}}{T} =
           \left(\frac{g_{*S}(T)}{g_{*S}(T_d)}\right)^{1/3}
\end{equation}
Since $g_{*S}(1 \mbox{MeV}) = 10.75$, the bound
(\ref{experiment}) implies
$g_{*S}(T_d) > 17.4$ which in turn forces $T_d > 200$ MeV.
This result applies when $m \ll 1$ MeV. When $m \sim 1$
MeV, Eqs. (\ref{phi role}) and (\ref{T phi}) have to be
modified but still one can restrict $g$ when
\begin{equation}
m \leq 2.6 \ \mbox{MeV}
\end{equation}
In this mass range we obtain the condition
\begin{equation} \label{rates}
\Gamma(T) < H(T) \ \ \mbox{for} \ 1 \ \mbox{MeV} \leq T
                                      \leq 100 \ \mbox{MeV}.
\end{equation}
It is now easy to extract a bound on $g$
using the condition (\ref{rates}) together with the
expressions for $\Gamma$
and $H$, Eqs. (\ref{Gamma}) and (\ref{Hubble}). For that
range of energies,
$Q = e$ dominates the $\phi$ interaction. We get
\begin{equation} \label{BBN g}
g < 2 \ 10^{-7} \ \mbox{GeV}^{-1}
\end{equation}
for $m<2.6$ MeV.
The region limited by BBN is shown in Fig. 2.


\section{Extensions}

In this section we will consider first the case that $\phi$ is a
scalar
particle, and second we will generalize the couplings in such a way
that they have the full $SU(2) \times U(1)$ gauge invariance.

The motivation to consider a particle $\phi$ coupled
to two photons was
that such particle could be detected in some of the
undergoing axion
experiments. In the previous sections, we have considered
a pseudoscalar
particle with a coupling as in Eq. (\ref{lag.pseu}).
We could as well
consider a scalar particle $\phi_s$, with mass $m_s$,
that would couple as
\begin{equation} \label{lag.scal}
{\cal L} = \frac{1}{4} \ g_s \ F^{\mu \nu} F_{\mu \nu} \ \phi_s
\end{equation}

One can show that all the limits we have for
the pseudoscalar coupling
$g$ hold exactly for the scalar coupling $g_s$. Our Fig. 2 showing
the regions allowed for $g$ can be used without changes for
$g_s$ ( and $m_s$ instead of $m$ ).
The considerations regarding $\phi$ as dark matter can also be
translated into identical statements for $\phi_s$, and similarly
in Fig. 6 we can interchange $g$ by $g_s$. At this point it is
instructive to know that the optical experiments \cite{Bakalov},
mentioned in Sec. 3, will be able to distinguish between
the effects of a scalar $\phi_s$ and a pseudoscalar $\phi$ boson
since they lead to different signatures.

Our second point is related to gauge invariance. The Lagrangian
(\ref{lag.pseu}), containing a dimension five operator,
represents an
effective interaction. New physics at an energy scale
$\Lambda \sim g^{-1}$
would lead to the low energy interaction expressed in
(\ref{lag.pseu}).
This scale is much larger than the weak scale,
$\Lambda \gg G_F^{-1/2}$.
For energies in between $\Lambda$ and $G_F^{-1/2}$, the effective
interaction should have the full $SU(2) \times U(1)$
gauge invariance.
Our purpose is to explore the consequences of the imposition of
gauge invariance. To proceed, we need to specify
the behavior of $\phi$
under the gauge symmetry. For simplicity, we take $\phi$ as a
$SU(2) \times U(1)$ singlet.

There are two types of operators that are interesting
for our purposes
and preserve $SU(2) \times U(1)$ gauge invariance.
In the form of pieces
of the effective Lagrangian they are
\begin{eqnarray} \label{lag.gauge}
{\cal L_{BB}} & = & \frac{1}{8} \ g_{BB} \
                    \varepsilon_{\mu \nu \alpha \beta}
                  B^{\mu \nu} B^{\alpha \beta} \ \phi \nonumber \\
{\cal L_{WW}} & = & \frac{1}{8} \ g_{WW} \
                    \varepsilon_{\mu \nu \alpha \beta}
                  \vec{W}^{\mu \nu} \vec{W}^{\alpha \beta} \ \phi
\end{eqnarray}
where
\begin{eqnarray}
B_{\mu \nu} & = & \partial_{\mu} B_{\nu} - \partial_{\nu} B_{\mu} \\
\vec{W}_{\mu \nu} & = & \partial_{\mu} \vec{W}_{\nu} -
                        \partial_{\nu} \vec{W}_{\mu} +
                    \frac{e}{\sin \theta_W} \vec{W}_{\mu}
                    \times \vec{W}_{\nu}
\end{eqnarray}
and
\begin{eqnarray}
B_{\mu}   & = & -\sin \theta_W Z_{\mu} + \cos \theta_W A_{\mu} \\
W^3_{\mu} & = & \ \ \ \cos \theta_W Z_{\mu} +
                      \sin \theta_W A_{\mu} \\
W^1_{\mu} & = & \frac{1}{\sqrt{2}} \left(W_{\mu} +
                                     W^{\dagger}_{\mu} \right) \\
W^2_{\mu} & = & \frac{i}{\sqrt{2}} \left(W_{\mu} -
                                     W^{\dagger}_{\mu} \right)
\end{eqnarray}
being $Z_{\mu}$, $A_{\mu}$ and $W_{\mu}$ the
fields corresponding to the particles $Z$, $\gamma$ and
$W^{\pm}$; $\theta_W$ is the weak mixing angle.
Both operators (\ref{lag.gauge})
lead to $\phi$ coupling to two photons exactly as in
Eq. (\ref{lag.pseu}). However they lead to more than this,
as we will now explore for each piece of the Lagrangian
(\ref{lag.gauge}) in turn.

${\cal L}_{BB}$ leads to the couplings $\gamma \gamma \phi$,
$Z Z \phi$ and $\gamma Z \phi$
\begin{eqnarray} \label{BB}
{\cal L_{BB}} & = &  \frac{1}{8} \ g_{BB} \cos^2 \theta_W
                         \ \varepsilon_{\mu \nu \alpha \beta}
                            F^{\mu \nu} F^{\alpha \beta} \ \phi
                            \nonumber \\
           & + &  \frac{1}{8} \ g_{BB} \sin^2 \theta_W
                         \ \varepsilon_{\mu \nu \alpha \beta}
                            Z^{\mu \nu} Z^{\alpha \beta} \ \phi
                            \nonumber \\
           & - & \frac{1}{4} \ g_{BB} \sin \theta_W \cos \theta_W
                         \ \varepsilon_{\mu \nu \alpha \beta}
                            F^{\mu \nu} Z^{\alpha \beta} \ \phi
\end{eqnarray}
where
$Z_{\mu \nu} = \partial_{\mu} Z_{\nu} - \partial_{\nu} Z_{\mu}$.

${\cal L}_{WW}$ is richer in structures
\begin{eqnarray} \label{WW}
{\cal L_{WW}} & = & \frac{1}{8} \ g_{WW} \sin^2 \theta_W
                         \ \varepsilon_{\mu \nu \alpha \beta}
                            F^{\mu \nu} F^{\alpha \beta} \ \phi
                            \nonumber \\
           & + & \frac{1}{8} \ g_{WW} \cos^2 \theta_W
                         \ \varepsilon_{\mu \nu \alpha \beta}
                            Z^{\mu \nu} Z^{\alpha \beta} \ \phi
                            \nonumber \\
           & + & \frac{1}{4} \ g_{WW} \sin \theta_W \cos \theta_W
                         \ \varepsilon_{\mu \nu \alpha \beta}
                            F^{\mu \nu} Z^{\alpha \beta} \ \phi
                            \nonumber \\
           & + & ...
\end{eqnarray}
namely, it generates similar couplings to
those generated by ${\cal L}_{BB}$
and in addition it generates $WW \phi$,
$WWZ \phi$ and $WW \gamma \phi$
couplings not explicitly written in the above expression.

{}From the phenomenological point of view the
most interesting interaction,
apart from $\gamma \gamma \phi$, is $\gamma Z \phi$.
It would contribute
to $e^+ e^- \rightarrow Z \rightarrow \gamma \phi$,
giving rise to single
photons with unbalanced momentum at the $Z$ peak.
Such a signature
has been found at LEP, consistent with the
Standard Model expectations,
$e^+ e^- \rightarrow Z \rightarrow \nu \bar{\nu} \gamma $.
The observed
signal is consistent with the Standard Model prediction
( $N_{\nu}=3$ ).
This implies limitations on the strength of
the different and a priori
independent pieces of the effective Lagrangian (\ref{lag.gauge}).
We will not allow for unnatural cancellations of the
effects produced by
different pieces, and thus we will consider one piece at a time.

Let us consider ${\cal L}_{BB}$. The observations made by the OPAL
collaboration at LEP \cite{OPAL} place a limit on the combination
\begin{equation}
g_{\gamma Z \phi} = -g_{BB} \ \sin \theta_W \cos \theta_W .
\end{equation}
Since we identify the $\gamma \gamma \phi$
coupling in Eq. (\ref{BB}) as
\begin{equation}
g = g_{BB} \ \cos^2 \theta_W
\end{equation}
we have that
\begin{equation}
g = -g_{\gamma Z \phi} \ \cot \theta_W
\end{equation}
and thus the experimental limit on $g_{\gamma Z \phi}$
translates into a limit on $g$. We get
\begin{equation} \label{BBlimit}
g \leq 1.2 \ 10^{-4} \ \mbox{GeV}^{-1}
\end{equation}
This limit holds provided $\phi$ does not decay inside de detector,
see Eqs. (\ref{length}) and (\ref{lifetime}), and
provided $m < 64$ GeV \cite{OPAL}.

A similar reasoning leads to a limit on $g$ when
considering ${\cal L}_{WW}$
\begin{equation} \label{WWlimit}
g \leq 3.6 \ 10^{-5} \ \mbox{GeV}^{-1} .
\end{equation}

We take as our result the {\bf less} restrictive bound
(\ref{BBlimit}). We
stress that our hypotheses are that $\phi$ is a
$SU(2) \times U(1)$ singlet
and that there are not unnatural cancellations
of the effects caused
by ${\cal L}_{BB}$ and ${\cal L}_{WW}$. As was discussed
in Sec. 2,
the coupling $g$ in the Lagrangian (\ref{lag.pseu})
is most restricted by
the ASP results: $g \leq 5.5 \ 10^{-4}$ GeV$^{-1}$.
In this section,
we have extended the Lagrangian (\ref{lag.pseu})
such that it possesses
gauge invariance. The appearance of a coupling
$\gamma Z \phi$ makes
possible to have a strongest bound:
$g \leq 1.2 \ 10^{-4}$ GeV$^{-1}$, using now the OPAL results.


\section{Summary and conclusions}

Most experiments searching for axions are based on its coupling to
two photons.
These experiments are also sensitive to a pseudoscalar
( or scalar particle ) $\phi$ that
couples only to two photons, and not to leptons, quarks
and nucleons. Motivated by this fact, we have examined the
constraints on such a particle, and
investigated to which extent $\phi$ can be the dark
matter of the Universe.
Some of the constraints can be deduced quite easily from
studies on the axion,
and other constraints have been deduced in this paper.

The laboratory, astrophysical and cosmological
limits are shown in Fig. 2.
High energy searches of $e^ + e^- \rightarrow \gamma + invisible$
give
the best constraints of what we have classified as
non-dedicated experiments.
Among the dedicated experiments, the solar flux detection
gives strong limits
once one assumes $\phi$ production in the solar core.
Laser experiments
give poorer limits, but are free of any astrophysical assumption.
The telescope
search give very strong constraints, but in a very
limited range of $\phi$
masses. Consideration of He burning stars allows
to place very stringent
limits for $m \leq O(10\ \mbox{keV})$. For higher $\phi$ masses,
one has to
rely on the limits from SN1987A observations and from
considerations of big bang nucleosynthesis.

We have studied the cosmological evolution of the $\phi$
species, and
calculated the relic $\phi$ density. The interesting
range of masses and
couplings that leads to a $\phi$ density such that
$\phi$ can be at least
the galactic dark matter is shown in Fig. 6.
Unfortunately, the mass
range interesting for dark matter is much higher
than the masses to which
most of the existing experiments are sensitive. Only the
telescope search
experiment is sensitive to masses that are close to the
dark matter range.

Another conclusion that we have reached has been that,
if $\phi$ is a relic species, it must be a hot relic.

The case that $\phi$ is a scalar particle is very similar to the
pseudoscalar case, regarding the
constraints on the coupling and our conclusions on dark matter.

A final aspect we have studied is
the $SU(2) \times U(1)$ gauge invariant
generalization of the $\phi$ interactions.
In addition to the vertex
$\gamma \gamma \phi$, one then has a vertex of the type
$\gamma Z \phi$, as well as other exotic couplings.
Experimental data from
$e^+ e^- \rightarrow \gamma + missing$ at the $Z$ peak
leads to limits on the coupling $g$, that are
stronger than the limits obtained from this process without
the gauge invariant generalization.


\section*{Acknowledgements}

We are grateful to Dr. G. Raffelt for useful comments on the
stellar energy loss arguments.

We thank the Theoretical Astroparticle Network for support under
the EEC
Contract No. CHRX-CT93-0120 ( Direction Generale 12 COMA ).
This work has been partially supported by the CICYT Research Project
Nos. AEN-93-0474 and AEN-93-0520.
R.T. acknowledges a FPI Grant from Ministeri d'Educaci\'{o}
i Ci\`{e}ncia (Spain).

\newpage

\newpage
\section*{Figure captions}
\bigskip

\noindent{\bf Figure 1:} (a) Diagram of quarkonium
decay into $\phi$
and a photon; (b) Diagram of $e^+e^-$ annihilation
into $\phi$ and a photon.

\medskip
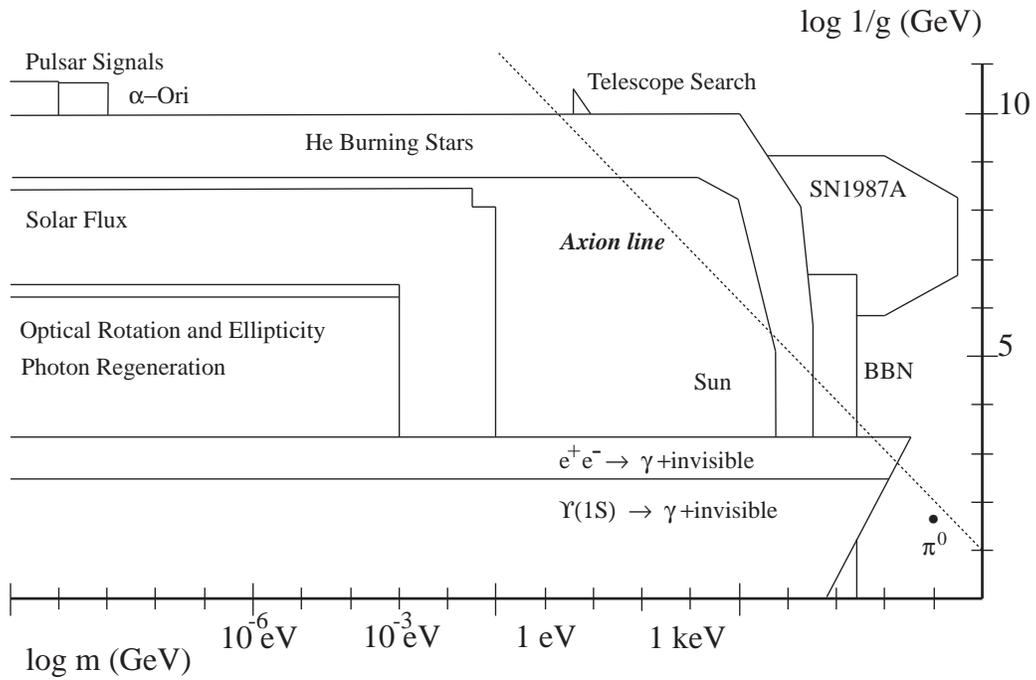
\noindent{\bf Figure 2:} Excluded regions for the mass
$m$ and coupling
$g$ of $\phi$, coming from laboratory, astrophysical
and cosmological
considerations. The line that relates coupling and mass
for the axion
is shown. We also show the coupling and mass of $\pi^0$.

\medskip
\noindent{\bf Figure 3:} Diagrams contributing to (a) rotation (b)
ellipticity and (c) photon regeneration effects of the
laser experiments.

\medskip
\noindent{\bf Figure 4:} (a) $\phi$ are produced in the Sun and
(b) reconverted in X-rays.

\medskip
\noindent{\bf Figure 5:} Processes contributing to $\phi$ emission
in the supernova core. In the second diagram, $N$ stands for $n$
or $p$. There are also similar diagrams not displayed where the
photon is attached to a initial $p$.

\medskip
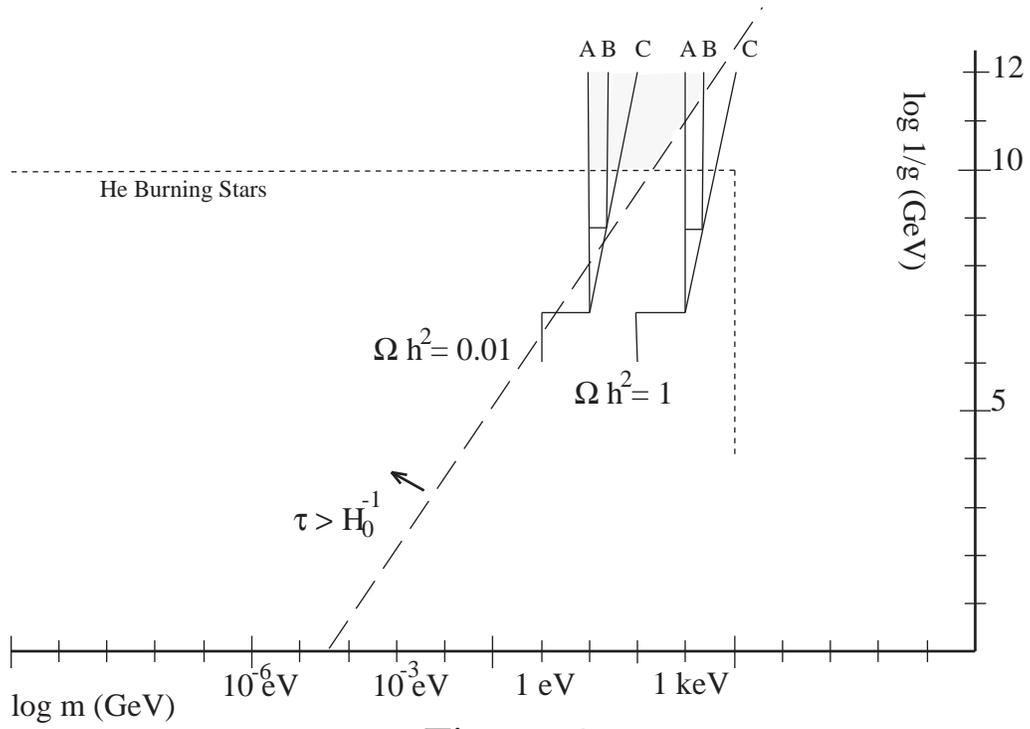
\noindent{\bf Figure 6:} The lines $\Omega_{\phi} h^2= 0.01$ and
$\Omega_{\phi} h^2= 1$ are represented as a
function of $g$ and $m$.
Each condition is calculated in three different cases, according
to how we extrapolate the degrees of freedom in
the early Universe
for $10^9 \geq T \geq 300$ GeV. In (A) there are no more excited
degrees of freedom, while in (C) they increase at the same rate
as a function of $T$ than they do for $T \leq 300$ GeV. In (B)
they increase at $T \sim 300$ GeV as in a SUSY theory and stay
constant at higher temperatures. The condition that $\phi$ has
survived until the present time, $\tau > H_0^{-1}$,
is also shown.


\begin{thebibliography}{99}

\bibitem{Peccei/Quinn} R.D. Peccei and H. Quinn,
      {\sl Phys. Rev. Lett.} {\bf 38}, 1440 (1977) and
      {\sl Phys. Rev.}  {\bf D16}, 1791 (1977).
\bibitem{Weinberg...} S. Weinberg,
      {\sl Phys. Rev. Lett.} {\bf 40}, 223 (1978);

      F. Wilczek,
      {\sl Phys. Rev. Lett.} {\bf 40}, 279 (1978).

      For invisible axions see:

      J.E. Kim,
      {\sl Phys. Rev. Lett.} {\bf 43}, 103 (1979);

                 M.A. Shifman, A.I. Vainshtein and V.I. Zhakharov,
      {\sl Nucl. Phys.} {\bf B166}, 493 (1980);

                 M. Dine, W. Fischler and M. Srednicki,
      {\sl Phys. Lett.} {\bf 104B}, 199 (1981);

                 A.R. Zhitnitskii,
      {\sl Sov. J. Nucl. Phys.} {\bf 31}, 260 (1980).
\bibitem{Peccei/Jarlskog} R.D. Peccei,
      in ``CP Violation'', ed. C. Jarlskog. World Scientific Publ.,
      pp. 503-551  1989.
\bibitem{Sikivie} P. Sikivie,
      in {\sl Proc. 17th John Hopkins Workshop on Particles and the
      Universe}, Budapest, pp. 139-151 1993.
\bibitem{Ball} Crystal Ball collaboration, D. Antreasyan {\sl et al.}
      {\sl Phys. Lett.} {\bf 251B}, 204 (1990).
\bibitem{Hearty}  ASP collaboration, C. Hearty {\sl et al.},
      {\sl Phys. Rev.} {\bf D39}, 3207 (1989);
\bibitem{Others}  CELLO collaboration, H.J. Behrend {\sl et al.},
      {\sl Phys. Lett.} {\bf B215}, 186 (1988);

      MAC collaboration, W.T. Ford {\sl et al.},
      {\sl Phys. Rev.} {\bf D33}, 3472 (1986);

      H. Wu, Ph.D. Thesis, Univ. Hamburg (1986);

      VENUS collaboration, K. Abe {\sl et al.},
      {\sl Phys. Lett.} {\bf B232}, 431 (1989);

      ALEPH collaboration, D. Buskulic {\sl et al.},
      {\sl Phys. Lett.} {\bf B313}, 520 (1993);

      L3 collaboration, O. Adriani {\sl et al.},
      {\sl Phys. Lett.} {\bf B292}, 463 (1992);

      L3 collaboration, B. Adeva {\sl et al.},
      {\sl Phys. Lett.} {\bf B275}, 209 (1992);

      OPAL collaboration, M.Z. Akrawy {\sl et al.},
      {\sl Z. Phys.} {\bf C50}, 373 (1991).

\bibitem{Maiani} L. Maiani, R. Petronzio and E. Zavattini,
      {\sl Phys. Lett.} {\bf B175}, 359 (1986).
\bibitem{VanBibber1} K. van Bibber {\sl et al.},
      {\sl Phys. Rev. Lett.} {\bf 59}, 759 (1987).
\bibitem{Cameron} R. Cameron {\sl et al.},
      {\sl Phys. Rev.} {\bf D47}, 3707 (1993).
\bibitem{Sikivie.idea} P. Sikivie,
      {\sl Phys. Rev. Lett.} {\bf 51}, 1415 (1983);

      P. Sikivie,
      {\sl Phys. Rev.} {\bf D32}, 2988 (1985).
\bibitem{VanBibber2} K. van Bibber, P.M. McIntyre, D.E. Morris
                     and G.G. Raffelt,
      {\sl Phys. Rev.} {\bf D39}, 2089 (1989).
\bibitem{Lazarus} D. Lazarus {\sl et al.},
      {\sl Phys. Rev. Lett.} {\bf 69}, 2333 (1992).
\bibitem{Kephart/Turner} W. Kephart and T.J. Weiler,
      {\sl Phys. Rev. Lett.} {\bf 58}, 171 (1987);

                         M.S. Turner,
      {\sl Phys. Rev. Lett.} {\bf 59}, 2489 (1987).
\bibitem{Bershady} M.A. Bershady, M.T. Ressell and M.S. Turner,
      {\sl Phys. Rev. Lett.} {\bf 66}, 1398 (1991).
\bibitem{Wuensch/Hagmann} S. De Panfilis {\sl et al.},
      {\sl Phys. Rev. Lett.} {\bf 59}, 839 (1987);

                          W.U. Wuensch {\sl et al.},
      {\sl Phys. Rev.} {\bf D40}, 3153 (1989);

                          C. Hagmann, P. Sikivie, N.S. Sullivan
                          and D.B. Tanner,
      {\sl Phys. Rev.} {\bf D42}, 1297 (1990).
\bibitem{Bakalov} D. Bakalov {\sl et al.},
      {\sl Nucl. Phys. ( Proc. Suppl. )} {\bf B35}, 180 (1994);

                  D. Bakalov {\sl et al.},
      presented at the MG7 meeting.
\bibitem{Raffelt} G.G. Raffelt,
      {\sl Phys. Rev.} {\bf D33}, 897 (1986).
\bibitem{Vorobyov} P.V. Vorobyov and I.V. Kolokolov,
      astro-ph-9501042.
\bibitem{Kolb} E.W. Kolb and M.S. Turner,
      {\sl The Early Universe}, Frontiers in Physics.
      Addison-Wesley Publish. Co., 1990.
\bibitem{Raffelt/Dearborn} G.G. Raffelt and D.S.P. Dearborn,
      {\sl Phys. Rev.} {\bf D36}, 2211 (1987).
\bibitem{Raffelt.PR} G.G. Raffelt,
      {\sl Phys. Rep.} {\bf 198}, 1 (1990).
\bibitem{Starkman/Salati} E.D. Carlson and P. Salati,
      {\sl Phys. Lett.} {\bf B218}, 79 (1989);

                   G.G. Raffelt and G.D. Starkman,
      {\sl Phys. Rev.} {\bf D40}, 942 (1989).
\bibitem{Schramm} D.N. Schramm,
      in {\sl Lepton and Photon Interactions}, proceedings of
      the International Symposium on Lepton and Photon Interactions
      at High Energies, Hamburg, Germany, 1987, edited by W. Bartel
      and R. R\"{u}ckl.

      {\sl Nucl. Phys.} B (Proc. Suppl.) {\bf 3}, 471 (1987).
\bibitem{Shapiro} S.L. Shapiro and S.A. Teukolsky,
      {\sl Black Holes, White Dwarfs and Neutron Stars},
      Wiley-Interscience. John Wiley \& Sons, 1983.
\bibitem{Mohanty} S. Mohanty and S.N. Nayak,
      {\sl Phys. Rev. Lett.} {\bf 70}, 4038 (1993).
\bibitem{Carlson} E.D. Carlson,
      {\sl Phys. Lett.} {\bf B344}, 245 (1995).
\bibitem{Olive} K.A. Olive, D.N. Schramm and G. Steigman,
      {\sl Nucl. Phys.} {\bf B180}, 497 (1981).
\bibitem{Walker} T.P. Walker, G. Steigman, D.N. Schramm,
                 K.A. Olive and H. Kang,
      {\sl Astrophys. J.} {\bf 376}, 51 (1991).
\bibitem{OPAL} OPAL collaboration, R. Akers {\sl et al},
      {\sl Z. Phys.} {\bf C65}, 47 (1995).

\end{thebibliography}
\end{document}